\documentclass{article}
\usepackage{amssymb,latexsym}
\usepackage[dvipdf]{epsfig}
\usepackage{color}
\newcommand{\be}{\begin{equation}}
\newcommand{\ee}{\end{equation}}
\newcommand{\no}{\noindent}
\newcommand{\n}{\label}
\newcommand{\ga}{\gamma}
\newcommand{\La}{\Lambda}
\usepackage{graphicx}
\begin{document}
\def\be{\begin{equation}}
\def\ee#1{\label{#1}\end{equation}}
\def\v{\varphi}

\title{Tachyonization of the $\La$CDM cosmological model}

\author{\emph{Luis P. Chimento \footnote{chimento@df.uba.ar}$\,$  and M\'onica Forte \footnote{forte.monica@gmail.com}}
\\
Departamento de F\'{\i}sica, Facultad de Ciencias Exactas y
Naturales,  \\Universidad de Buenos Aires, Ciudad
Universitaria, \\Pabell\'on I, 1428 Buenos Aires, Argentina\\
\emph{Gilberto M. Kremer \footnote{kremer@fisica.ufpr.br}}\\
 Departamento de F\'{\i}sica,
Universidade Federal do Paran\'a,\\ Caixa Postal 19044, 81531-990 Curitiba, Brazil\\
 \emph{Marlos O. Ribas \footnote{marlos@fisica.ufpr.br}}\\
 Faculdades Integradas Esp\'{\i}rita, Rua Tobias de
Macedo Jr.  333,\\ 82010-340 Curitiba, Brazil}

\date{}
\maketitle

\begin{abstract}
 In this work a tachyonization of the $\Lambda$CDM model for a spatially flat Friedmann-Robertson-Walker space-time is proposed.  A tachyon field  and a cosmological constant are considered as the sources of the gravitational field. Starting from a  stability analysis and from the exact solutions for a standard tachyon field driven by a given potential, the search for a large set of cosmological models which contain the $\Lambda$CDM model   is investigated. By the use of internal transformations two new kinds of tachyon fields are derived from the standard tachyon field, namely, a complementary and a phantom tachyon fields. Numerical solutions for  the three kinds of tachyon fields are determined and it is shown that the standard and complementary tachyon fields reproduces the $\Lambda$CDM model as a limiting case. The standard tachyon field can also describe a transition from an accelerated to a decelerated regime, behaving as an inflaton field at early times and as a matter field at late times. The complementary tachyon field always behaves as a matter field. The phantom tachyon field is characterized by a rapid expansion where its energy density increases with time.
\end{abstract}

\section{Introduction}

Nowadays  there    exists  consensus  that  at  early  times  an  inflaton  field  has  driven  the Universe to a rapid accelerated expansion. This period was followed by a decelerated era dominated by a matter field and then, by another accelerated period dominated by dark energy. Several cosmological models were proposed to  do  the job, based on rolling tachyon field from string theory. Also, much effort has been devoted to the study of tachyon potential and classical solutions on an unstable D-brane system. Actually, the energy-momentum tensor of the tachyon field can be seen as a sum of two tensors, associated with dark matter and vacuum energy density respectively \cite{plb}, whereas the tachyon potential has an unstable maximum at the origin decaying to almost zero as the field goes to infinity.  Depending on this asymptotic behavior several works have been carried out on tachyonic  dark energy, \cite{Bagla:2002yn}-\cite{Copeland:2006wr}, on tachyonic  dark matter, \cite{DeBenedictis:2004wj}-\cite{Padmanabhan:2002cp} and inflation models \cite{Sami:2002zy},\cite{Paul:2005xh}.
Also, in Refs. \cite{Chakraborty:2008uj}-\cite{Herrera:2004dh} it was assumed an interaction for good fit to the Supernovae and CMB data.

Although in  k-essence cosmologies,  the stability of the k-essence with respect to small wavelength perturbations requires a positive sound speed $c_s^2={\dot p/\dot \rho}$, in Ref. \cite{car} it was shown that a positive
sound speed is not a sufficient condition for the theory to be stable. Hence, non standard tachyon fields have been generated assuming a sound speed proportional to that of the standard tachyon field in \cite{LPC} for FRW and in \cite{Chimento:2005ua} for Bianchi type I cosmologies.

In this work it is analyzed the evolution of a Universe whose gravitational sources are a cosmological constant and a tachyon field, and its aim is to determine a large set of cosmological models which contain the $\Lambda$CDM model as a special case. Usually in the $\Lambda$CDM cosmological model it is assumed that the pressureless
component includes all types of matter known from laboratory experiments
(protons, neutrons, photons, neutrinos, etc.) as well as non-relativistic non-baryonic
cold dark matter (whose energy-momentum tensor is dust-like in the first approximation $0 < p <<\rho$).
This is an approximation only justified by the simplicity of the model. However, we think that
it would be better to consider the possibility of including a large set of nearly
pressureless components. Consequently, we have introduced the tachyon field as
representative of the above variety of components because it may behave as dust like
in first approximation. In fact, the potential we have chosen for the tachyon field leads to
an effective equation of state which interpolates between a nearly dust era at early times
and a de Sitter stage at late times. In this way we are considering a set of enlarged
$\Lambda$CDM cosmological model with the possibility of finding more realistic
framework for the present universe.
The search for this large set of cosmological models makes use of the stability analysis of the standard tachyon field and internal transformations. In the Ref. \cite{LPC} one of the authors introduced the extended tachyon fields. These enlarged set of fields  can be split into three classes, so apart from the class containing the standard tachyon field two new classes arise, namely, a complementary and a phantom tachyon fields. The three classes of tachyon fields will be  analyzed and shown that the limiting cases of the standard and complementary tachyon fields tend to the $\Lambda$CDM model whereas the standard tachyon field is able to describe the inflationary period.
We think that any model obtained from a symmetry transformation of the dynamical
equations should be taking into account as a realizable model. From this point of view
it is interesting also to investigate the phantom tachyon field.

The work is organized as follows: in Section II a stability analysis of the standard tachyon field is performed in order to determine the asymptotically stable solutions. The tachyonization process is the subject of Section III where the complementary and the phantom tachyon fields are introduced. In Section IV a numerical analysis of the three kinds of tachyon fields is considered and a discussion on the regimes is carried out. Finally in Section V it is given a summary of the main conclusions.

\section{Stability Analysis}

Let us present the Lagrangian density for the tachyon field $\v$
\be
\n{l}
{\cal L}_\v= -V(\v)\sqrt{1-\partial_\mu\v\partial^\mu\v},
\ee
where $V(\v)$ denotes the potential of the tachyon field.
One can associate an energy-momentum tensor for the tachyon field so that its energy density $\rho_\v$
and pressure $p_\v$ -- in a homogeneous and isotropic space-time represented by the spatially flat Friedmann-Robertson-Walker Universe -- are given by
\be
\rho_\v={V\over \sqrt{1-\dot\v^2}},\qquad
p_\v=-{V\sqrt{1-\dot\v^2}}.
\ee{3}

Let us consider the gravitational field generated by
a cosmological constant $\Lambda$ and  a tachyon field $\v(t)$ with energy density
$\rho_\v$. In this case the Friedmann and conservation equations become
\be
3H^2=\Lambda+\rho_\v,
\ee{1}
\be
\dot \rho_\v+3H(\rho_\v+p_\v)=0.
\ee{4}
where $H=\dot a(t)/a(t)$ is the Hubble parameter and $a(t)$ is the cosmic scale factor.
The evolution equation for the tachyon field  that follows from Eqs. (\ref{3}) and (\ref{4}) is
\be
{\ddot\v }+(1-\dot\v^2)\left[3H\dot\v+{1\over V}{dV\over d\v}\right]=0.
\ee{2}
Note that the dynamic of the tachyon field driven by the potentials $V$ and $V\to V_0\,V$ is the same for any multiplicative constant $V_0$.

In order to analyze the stability of  solutions let us relate the pressure and the energy density
 of the tachyon field by a barotropic index $\gamma$, so that $p_\v=(\gamma-1)\rho_\v$.  In this case it is easy to
 obtain from Eqs.(\ref{3}) that $\dot\v^2=\gamma$, whereas from  eqs. (\ref{1}) and (\ref{4}) it follows that
 \be
 3H^2=\Lambda+\frac{V}{\sqrt{1-\gamma}},
 \ee{61}
\be
\dot\rho_\v+\frac{3\gamma HV}{\sqrt{1-\gamma}}=0
 \ee{5}
 Now the differentiation of Eq. (\ref{61}) with respect to time and the use of Eq. (\ref{5})
leads to the differential equation for the barotropix index
\be
 \dot\gamma=2(\gamma-1)\left(3H\gamma+{\dot V\over V}\right).
 \ee{6}

 To obtain solutions that are asymptotically stable, the barotropic index should tend to a constant value, namely, $\gamma=\gamma_0$. In this case one can get from (\ref{6}) the asymptotically differential equation
 \be
 {\dot V\over V}\simeq-3\gamma_0{\dot a\over a}, \qquad\hbox{so that}\qquad V\simeq{V_0\over a^{3\gamma_0}}.
 \ee{7}
 From the knowledge of the asymptotic value for the potential density it follows the corresponding asymptotic value for the energy density, namely,
 \be
 \rho_\v={V_0\over\sqrt{1-\gamma_0}}\,\, a^{-3\gamma_0}.
 \ee{8}
 Equation (\ref{6}) can be rewritten thanks to ({\ref{7})  as
 \be
 \dot\gamma=6H(\gamma-1)(\gamma-\gamma_0).
 \ee{9}
The solution of the differential equation (\ref{9}) for the case where $\gamma_0\neq1$ is
\be
\gamma={\gamma_0+c_1a^{6(\gamma_0-1)}\over 1+c_1a^{6(\gamma_0-1)}}
\ee{10}
where $c_1$ is an integration constant. One can infer from the above equation that $\gamma$ tends asymptotically
 to $\gamma_0$ once $\gamma_0<1$. On the other hand, when $\gamma_0=1$ the solution of (\ref{9}) reads
 \be
 \gamma=1-{1\over c_2+6\ln a},
 \ee{11}
with $c_2$ denoting an integration constant. According to the last equation the barotropic index
tends to $\gamma_0=1$.

\section{The tachyonization process}

Motivated by the stability analysis of the previous section -- where the relationship (\ref{8}) between the energy density of the tachyon field and the cosmic scale factor is similar to that of a perfect fluid with constant barotropic index -- one is tempted to find and investigate exact solutions for tachyon fields driven by the following potential
 \be
 V(\v)=\frac{\Lambda\sqrt{1-\ga_0}}{\sinh^2{\frac{\sqrt{3\ga_0\Lambda}}{2}}\,\v},
  \ee{16}
where $\ga_0$ is a parameter of the model.
Although the potential diverges at early times, where $\v\to 0$, it
reasonably mimics the behavior of a typical potential in the condensate of
bosonic string theory.
One expects the potential to have a unique local
maximum at the origin and a unique global minimum away from the origin at
which $V$ vanishes \cite{ku}. In the most interesting case the global minimum is taken
to lie at infinity \cite{gibb}. Obviously more complicated potentials having the same limit could be analyzed
but we restrict to that of Eq. (\ref{16}). For this potential the tachyon field equation (\ref{2}) becomes
  \be
{\ddot\v }+(1-\dot\v^2)\left[3H\dot\v-\sqrt{3\ga_0\La}\coth{\frac{\sqrt{3\ga_0\La}}{2}\v}\right]=0.
\ee{tfe}

Exact solutions for the tachyon field were found using a linear field in several references \cite{alex,fein}. Also, this assumption has proved to be useful in the context of k-essence theories  \cite{stein},\cite{LPC}. With this idea in the mind, our purpose is to find an exact solution of the field equation (\ref{tfe}) by assuming a linear dependence of the tachyon field with the cosmological time,
 \be
 \v=\sqrt{b}\, t,       \qquad \dot\v^2=b,
 \ee{15}
where $0<b<1$ is a constant to be determined later on. The latter will be solution of  Eq. (\ref{tfe}) if
\be
H=\sqrt{\frac{\ga_0\La}{3b}}\coth{\sqrt{3\ga_0 b\La}\over2}\,t,
\ee{hb}
which, after integration we obtain the following cosmic scale factor
\be
a=a_0\left[\sinh{\frac{\sqrt{3\ga_0 b\La}}{2}\,t}\right]^{2/3b}.
\ee{sf}
By imposing the consistency of this cosmic scale factor -- expressed in terms of the linear tachyon field (\ref{15}) -- with respect to the solutions of the Friedmann equation (\ref{61}) for the potential (\ref{16}), namely,
\be
3H^2=\La\left[1+\frac{\sqrt{1-\ga_0}}{\sqrt{1-\dot\v^2}\,\sinh^2{\frac{\sqrt{3\ga_0 b\Lambda}}{2}}\,t}\right],
 \ee{00}
we get $b=\ga_0$. Therefore, for the linear exact solutions, the  energy density of the tachyon field is given by that of a perfect fluid, i.e., $ \rho_\v=C/a^{3\ga_0 }$, where $C=\La a_0^{3\ga_0 }$ is a constant. In this case the Friedmann's equation (\ref{00}) reads
 \be
 3H^2=\Lambda+{C\over a^{3\ga_0 }},
 \ee{13}
 whose integration leads to the knowledge of the cosmic scale factor as function of time:
 \be
 a=\left[\sqrt{C\over\Lambda}\sinh\left({\sqrt{3\ga_0^2\Lambda}\over2}\,t\right)\right]^{2\over3\ga_0}.
 \ee{14}
Furthermore, the final energy density and pressure of the tachyon field are given by
\be
\rho_\v=\frac{\La\sqrt{1-\ga_0}}{\sqrt{1-\dot\v^2}\,\sinh^2{\frac{\sqrt{3\ga_0\Lambda}}{2}}\,\v},
\ee{r1}
\be
p_\v=-{\La\sqrt{(1-\ga_0)(1-\dot\v^2)}\over\sinh^2{{\frac{\sqrt{3\ga_0\Lambda}}{2}}\,\v}},
\ee{p1}
respectively.
Clearly, in the limit $\ga_0\to 1$, this cosmological model -- whose sources include the cosmological constant and a tachyon field driven by the potential (\ref{16}) -- reduces to the standard $\La$CDM model. However, in the general case by considering other solutions different from the linear tachyon field, the model will have slightly dissimilar characteristics with respect to the $\La$CDM model. In the next section we shall investigate these kind of models by performing  a numerical analysis of their solutions.

To sum up, the tachyonization of the $\La$CDM model we have presented may considered as a process of constructing a large set of cosmological models, generated by a tachyon field, which contain the $\La$CDM model. It
means that the particular set of solution, corresponding the linear tachyon field solution along with the limit $\ga_0\to 1$, is the general exact cosmic scale factor solution of the $\La$CDM model.
In addition, for the case where the tachyon energy density has the perfect fluid form $\rho_\v=C/a^{3\ga_0}$, with $\ga_0\ne 1$, the time evolution of the cosmic scale factor is given by the solution  (\ref{14}) of the Friedmann equation (\ref{13}) and the tachyon field solution has the linear form (\ref{15}) with $b=\ga_0$. This represent the tachyonization of the flat FRW universe with cosmological constant filled with a perfect fluid.

Finally we comment that in the case without the cosmological constant, i.e., the case of the FRW universe filled with a perfect fluid, the potential  and the cosmic scale factor  reduce to
\be
V=\frac{4\sqrt{1-\ga_0}}{3\ga_0\phi^2},
\ee{va}
\be
a=\left[\frac{\sqrt{3\ga_0^2 C}}{2}\,t\right]^{2\over3\ga_0},
\ee{va1}
 after taking the limit $\La\to 0$ into the expressions (\ref{16}) and (\ref{14}) along with $\phi=\sqrt{\ga_0}\,t$. These results are exactly and the same found in Ref. \cite{alex,fein}. However, here we have shown that the potential (\ref{va}) gives the tachyonization of the FRW cosmological model with a perfect fluid. Hence, the remaining solutions of the tachyon field, corresponding to the solutions which differ from  $\phi=\sqrt{\ga_0}\,t$ are associated with a more general model containing it.

\subsection{Extended tachyon fields}

In a recent paper~\cite{LPC}  one of the authors introduced two new kinds of extended tachyon fields, apart from the well known tachyon field which was analyzed above. The usual tachyon field, with $\gamma<1$, can be associated with $0<\ga_0<1$. Below, we use the extended tachyon fields to cover all remaining values of $\ga_0$. The tachyonization of the model for $1<\ga_0$   will be achieved by the complementary  tachyon field $\v_c$, whereas for
$\ga_0<0$ by the phantom tachyon field $\v_{ph}$.  In Ref. \cite{acl} the phantom tachyon field was obtained by applying the dual symmetry transformation on the standard tachyon, generating an extended super accelerated tachyon field.

These two kinds of tachyon fields can be introduced from the tachyon field analyzed above by applying internal transformations as follows.

\vskip .5cm

\no {\it The complementary tachyon field } $\v_c$ is characterized by $1<\ga_0$ or $1<\dot\v_c^2$ and it can be generated from the standard tachyon field by making an internal transformation. In fact, a real tachyon field $\v_c$ is obtained from the energy density (\ref{r1}) and pressure (\ref{p1}) by replacing simultaneously $1-\ga_0\to -(1-\ga_0)$ and $1-\dot\v^2\to -(1-\dot\v_c^2)$, resulting
 \be
\rho_c={\Lambda{\sqrt{\ga_0-1}}\over \sqrt{\dot\v_c^2-1}\sinh^2{{\frac{\sqrt{3\ga_0\Lambda}}{2}}\,\v_c}},
 \ee{25}
\be
p_c={\Lambda{\sqrt{(\ga_0-1)(\dot\v_c^2-1)}}\over\sinh^2{{\frac{\sqrt{3\ga_0\Lambda}}{2}}\,\v_c}},
\ee{23}
\be
 {V_c={\Lambda\sqrt{\ga_0-1}\over\sinh^2{{\frac{\sqrt{3\ga_0\Lambda}}{2}}\,\v_c}}},
 \ee{24}
  \be
{\ddot\v_c }+(1-\dot\v_c^2)\left[3H\dot\v_c-\sqrt{3\ga_0\La}\coth{\frac{\sqrt{3\ga_0\La}}{2}\v_c}\right]=0.
\ee{18'}

\vskip .5cm

\no {\it The phantom tachyon field} $\v_{ph}$ is characterized by $\ga_0<0$, $\dot\v_{ph}^2=-\ga_0$.  It is generated  also from of the standard tachyon field by making an internal transformation. In this case the replacement $\sqrt{\ga_0}\to-\imath\sqrt{-\ga_0}$ and $\v\to\imath\v_{ph}$ in the equations (\ref{r1})-(\ref{p1}) gives
 \be
\rho_{ph}={\Lambda\sqrt{1-\ga_0}\over\sqrt{1+\dot\v_{ph}^2}\sinh^2{{\frac{\sqrt{-3\ga_0\Lambda}}{2}}\,\v_{ph}}},
 \ee{29}
\be
p_{ph}=-{\Lambda{\sqrt{(1-\ga_0)(1+\dot\v_{ph}^2)}}\over\sinh^2{{\frac{\sqrt{-3\ga_0\Lambda}}{2}}\,\v_{ph}}},
\ee{27}
 \be
 V_{ph}={{\Lambda\sqrt{1-\ga_0}}\over\sinh^2{{\frac{\sqrt{-3\ga_0\Lambda}}{2}}\,\v_{ph}}},
 \ee{28}
\be
{\ddot\v_{ph} }+(1+\dot\v_{ph}^2)\left[3H\dot\v_{ph}-\sqrt{-3\ga_0\La}\coth{\frac{\sqrt{-3\ga_0\La}}{2}\v_{ph}}\right]=0.
\ee{26}

We note that the cosmological expressions for the two new kinds of tachyon fields were found by using simple internal symmetries. In conclusion, we need all of them to describe the time evolution of the cosmic scale factor (\ref{14}) for all $\ga_0$ values. This close the tachyonization of the flat FRW universe with cosmological constant filled with a perfect fluid.

\section{Analysis for given potential  density}
 Motivated by the above results  one is also tempted to find more general cosmological solutions
 by specifying the tachyon potential and allowing for more general solutions of the tachyon field equation (\ref{tfe}) other than the linear one given by Eq. (\ref{15}). In those cases, the tachyon energy density has a more complicated dependence with the cosmic scale factor than the perfect fluid one $\rho_\v=C/a^{3\ga_0}$. To find the solutions  one has to solve the following coupled system of differential equations:

\vskip .5cm

\no {\it {\bf 1)} Standard tachyon field} $\v$

\be
 3H^2=\Lambda+{V(\v)\over \sqrt{1-\dot\v^2}},
 \ee{30}
 \be
{\ddot\v }+(1-\dot\v^2)\left(3H\dot\v+{1\over V}{dV\over d\v}\right)=0,
\ee{31}

\vskip .5cm

\no {\it  {\bf 2)} Complementary tachyon field} $\v_c$

\be
 3H^2=\Lambda+{V_c(\v_c)\over \sqrt{\dot\v_c^2-1}},
 \ee{32}
\be
{\ddot\v_c }-(\dot\v_c^2-1)\left(3H\dot\v+{1\over V_c}{dV_c\over d\v}\right)=0,
\ee{33}

\vskip .5cm

\no {\it  {\bf 3)} Phantom tachyon field} $\v_{ph}$

\be
 3H^2=\Lambda+{V_{ph}(\v_{ph})\over \sqrt{1+\dot\v_{ph}^2}},
 \ee{34}
\be
{\ddot\v_{ph} }+(1+\dot\v_{ph}^2)\left(3H\dot\v_{ph}-{1\over V_{ph}}{dV_{ph}\over d\v_{ph}}\right)=0.
\ee{35}
 In the above equations the potential densities $V(\v)$, $V_c(\v_c)$ and $V_{ph}(\v_{ph})$ are given by eqs. (\ref{16}), (\ref{24}) and (\ref{28}), respectively.

To find exact solutions of the  three systems of differential equations  is a very hard task and in the following numerical solutions of the three systems are analyzed.

In figures 1 to 5 it is plotted the numerical solutions of the above systems of differential equations for the following initial conditions $a(1)=1$, $\v(1)=\v_c(1)=\v_{ph}(1)=1$ and $\rho(1)=\rho_c(1)=\rho_{ph}(1)=1$. Moreover,
in all three cases the cosmological constant was taken  equal to $\Lambda=0.001$. Figures 1 to 4 refers to the standard and complementary tachyon fields whereas figure 5 correspond to the phantom one.
\begin{figure}
 \begin{center}
 \vskip1cm
 \includegraphics[height=5cm,width=7.5cm]{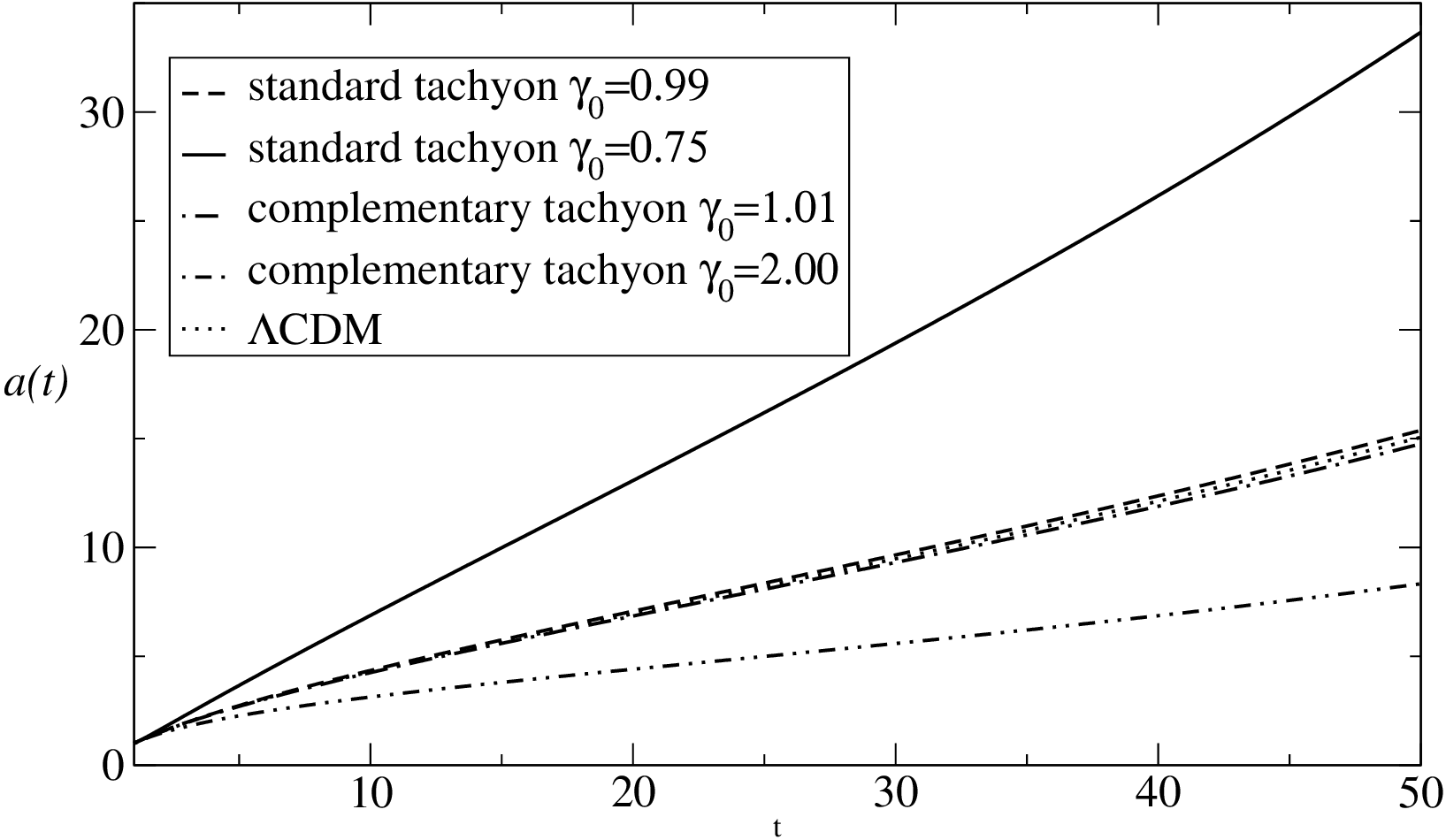}
  \caption{Cosmic scale factors as functions of time for standard ($\ga_0=0.99$  and $\ga_0=0.75$) and complementary ($\ga_0=1.01$  and $\ga_0=2$) tachyon fields. }
 \end{center}
 \end{figure}

 \begin{figure}
 \begin{center}
 \vskip1cm
 \includegraphics[height=5cm,width=7.5cm]{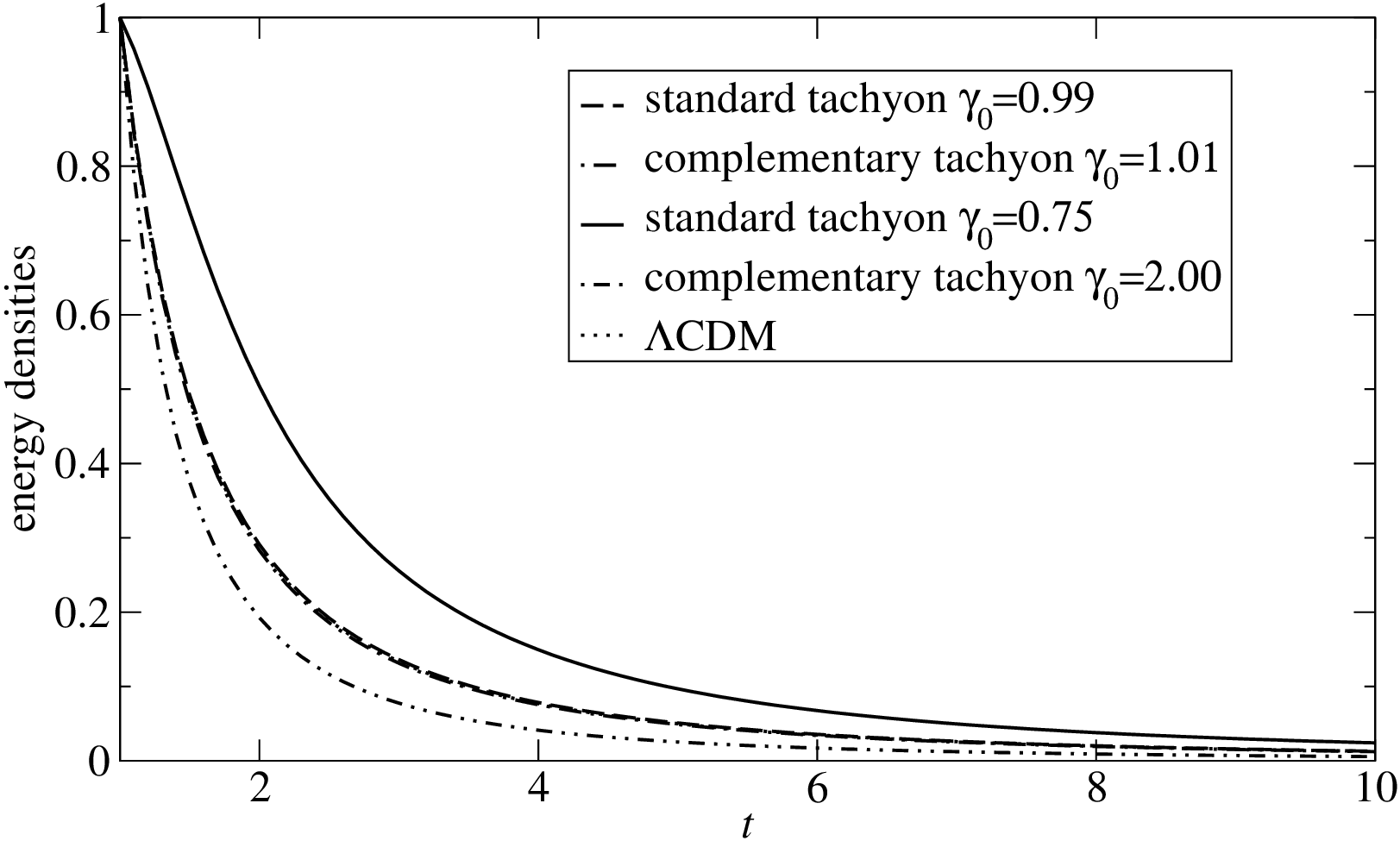}
  \caption{Energy densities as functions of time for standard ($\ga_0=0.99$  and $\ga_0=0.75$) and complementary ($\ga_0=1.01$  and $\ga_0=2$) tachyon fields.}
 \end{center}
 \end{figure}

 \begin{figure}
 \begin{center}
 \includegraphics[height=5cm,width=7.5cm]{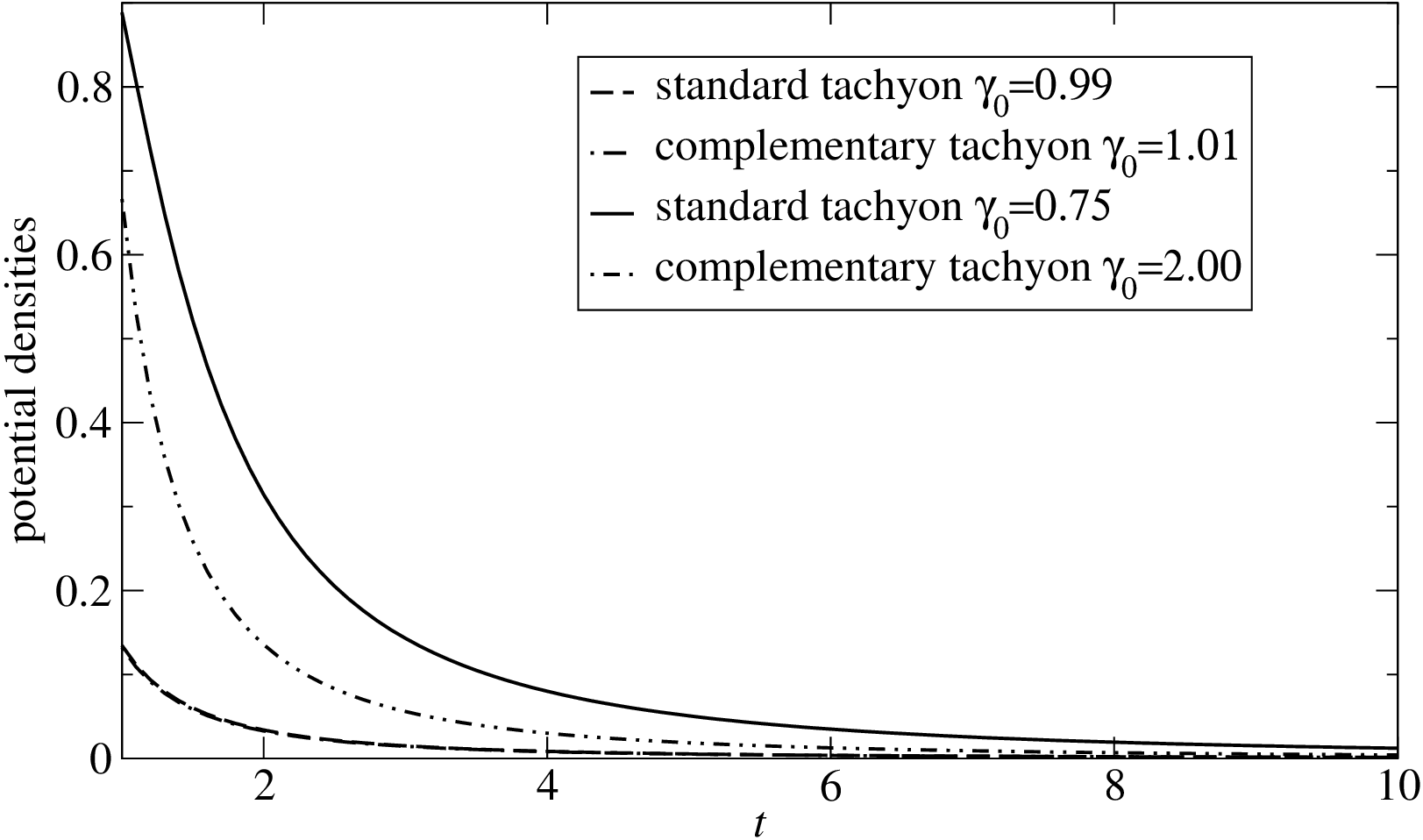}
  \caption{Potential densities as functions of time for standard ($\ga_0=0.99$  and $\ga_0=0.75$) and complementary ($\ga_0=1.01$  and $\ga_0=2$) tachyon fields.}
 \end{center}
 \end{figure}

\begin{figure}
 \begin{center}
 \includegraphics[height=5cm,width=7.5cm]{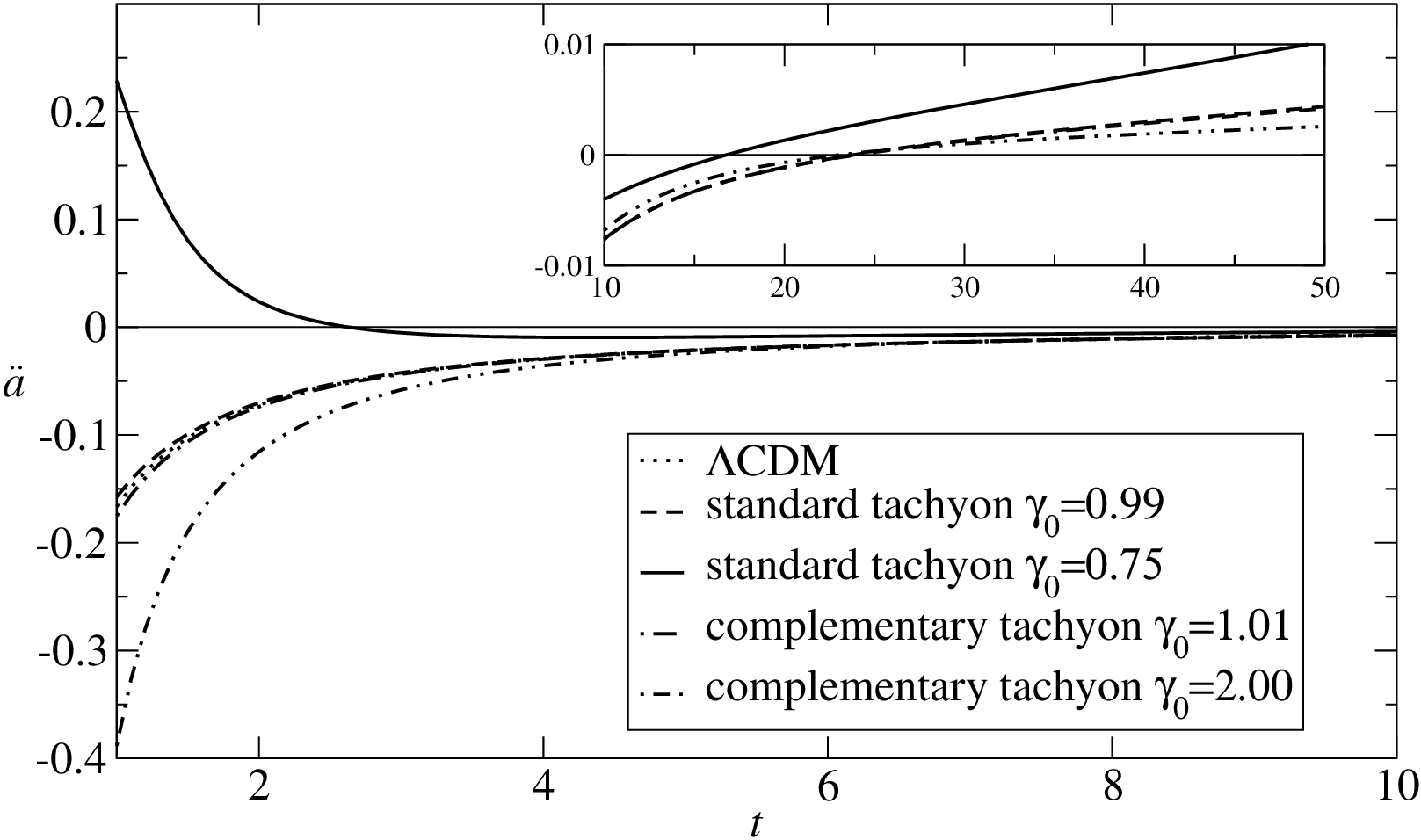}
  \caption{Accelerations as functions of time for standard ($\ga_0=0.99$  and $\ga_0=0.75$) and complementary ($\ga_0=1.01$  and $\ga_0=2$) tachyon fields.}
 \end{center}
 \end{figure}

 \begin{figure}
 \begin{center}
 \vskip1cm
 \includegraphics[height=5cm,width=7.5cm]{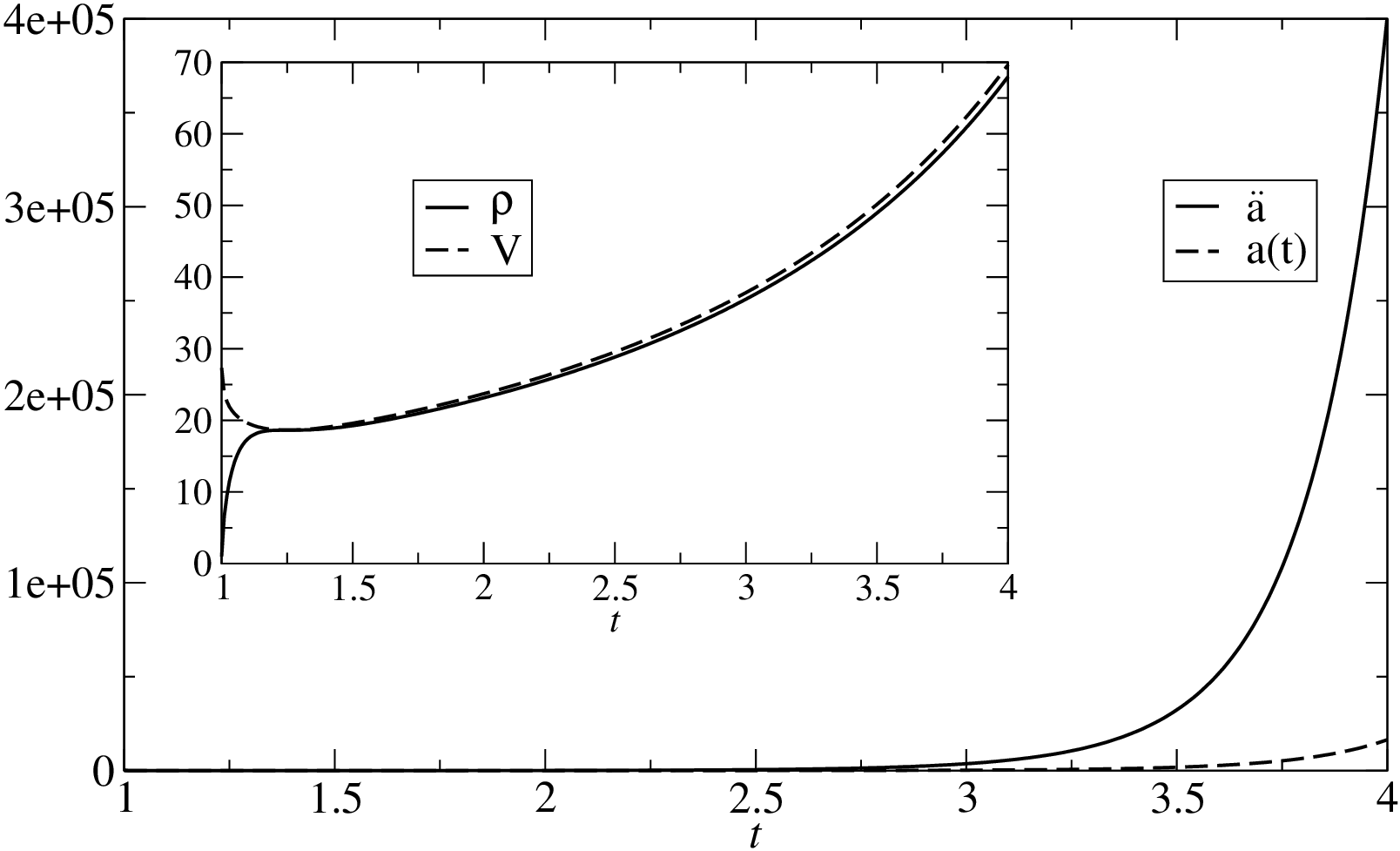}
  \caption{Cosmic scale factor, acceleration, energy density and potential density as functions of time for phantom tachyon field, for $\ga_0=-0.05$.}
 \end{center}
 \end{figure}

Let us  analyze the standard and complementary tachyon fields for values of $\ga_0\approx 1$ which are very closed to the $\Lambda$CDM model. For the standard tachyon field it was chosen $\ga_0=0.99$ whereas for the complementary  $\ga_0=1.01$.
For these two values one infers from figures 1 to 4 that there is no sensible difference of the solutions concerning the time evolution of the cosmic scale factors (fig. 1), energy densities (fig. 2), potential densities (fig. 3) and accelerations (fig.4) with the $\Lambda$CDM model. From these figures one concludes that the cosmic scale factors increase with time, the energy densities and the potential densities of the tachyon fields decrease with time, whereas the accelerations show a transition from a decelerated to an accelerated regime. The two tachyon fields here behave as
 matter fields which are responsible for the decelerated regime.

 The standard tachyon field with $\ga_0=0.75$ plays the role of an inflaton field which decays into a matter field. This can be observed from the behavior of its acceleration field in fig. 4, since at early times it begins with an accelerated regime which goes into a decelerated era returning afterwards to another accelerated regime where the cosmological constant predominates.  This kind of  behavior of the acceleration field was also obtained in the work~\cite{KA} where the potential density of the tachyon field was considered as an exponential type.
 The analysis of figures 1 to 3 for $\ga_0=0.75$ shows that the evolution with time of the cosmic scale factor is much more accentuated than that for the $\Lambda$CDM model, the energy  density decreases more slowly whereas the potential density is larger than the one for the $\Lambda$CDM model.

 The complementary tachyon field with $\ga_0=2$ represents stiff matter with a deceleration larger than the one corresponding to the $\Lambda$CDM model (see fig. 4). The cosmological constant is the responsible for the transition from a decelerated to accelerated regime. From figures 1 and 2 one can infer that for $\ga_0=2$ the increase with time of the cosmic scale factor is less accentuated than the one for the $\Lambda$CDM model, whereas the energy density decreases
 more rapidly. Furthermore, according to fig.3 the potential density stands between the potential densities of the standard tachyon field  with $\ga_0=0.75$ and the $\Lambda$CDM model.

 The phantom tachyon field has the typical characteristics of a phantom field, since its cosmic scale factor and acceleration field increase very rapidly with time (see fig. 5). Moreover, its energy density grows with time and the potential energy dominates for large times so that the energy density becomes equal to the potential density.

\section{Conclusions}

The tachyonization process we have introduced allows to think the $\Lambda$CDM model as included in a more general set of cosmological models with cosmological constant and containing a tachyon field driven by the potential (\ref{16}). This is a very interesting procedure to investigate models bearing certain resemblance to the $\Lambda$CDM model. The same process was applied to the FRW universe without cosmological constant and filled with a perfect fluid, taking the limit $\La\to 0$ of the above $\Lambda$CDM model. There we have obtained the standard inverse square potential, which has been extensively used for tachyon and k-essence cosmologies, and the corresponding power law solutions. The tachyonization of the $\Lambda$CDM model was determined from the stability analysis and from exact solutions of the standard tachyon field driven by the given potential (\ref{16}). From the use of simply internal transformations  two new kinds of extended tachyon fields were derived from the standard tachyon field: the  complementary and the phantom tachyon fields.

From the analysis of the numerical solutions for the three tachyon fields shown in figures 1 to 5 one concludes that: (i)  the standard and complementary  tachyon fields with $\gamma_0\approx1$ reproduce the $\Lambda$CDM model as a limiting case, since the time evolution of the cosmic scale factors (Fig. 1), energy densities (Fig.2) and accelerations (Fig. 4) are very close to each other; (ii) for values of $ \gamma_0<1$ the standard tachyon field  behaves  as an inflaton field at early times and as a matter field at late times describing a transition from an accelerated to a decelerated regime (Fig. 4). In this case the time evolution of the cosmic scale factor (Fig. 1) increases more accentuated than the $\Lambda$CDM model whereas its energy density (Fig. 2) decreases more slowly; (iii) for values of $ \gamma_0>1$ the complementary tachyon field  behaves as a decelerated matter field. Its energy density (Fig. 2) decays more rapidly with time in comparison to the  $\Lambda$CDM model, its cosmic scale factor (Fig. 1) increases more slowly and its deceleration is more accentuated (Fig. 4); (iii)
 the phantom tachyon field (Fig. 5) is characterized by a rapid accelerated expansion where its energy density increases with time.

\section*{Acknowledgments}

The authors acknowledge the partial
support under project 24/07 of the  agreement SECYT (Argentina) and CAPES 117/07 (Brazil).
LPC thanks the University of Buenos Aires for partial support under
project X224, and the Consejo Nacional de Investigaciones
Cient\'{\i}ficas y T\'ecnicas under project 5169. GMK acknowledges the support by
Conselho Nacional de Desenvolvimento Cient\'{\i}fico e Tecnol\'ogico (CNPq).



 \end{document}